\documentstyle[12pt,a4wide]{article}

\begin{document}
\begin{titlepage}

\begin{flushright}
UAB-FT-341\\
June 1994
\end{flushright}

\vspace{\fill}

\begin{center}
        {\LARGE \bf THE LAGRANGIAN LOOP REPRESENTATION
\vskip.3cm
OF LATTICE U(1) GAUGE THEORY}
\end{center}

\vspace{\fill}

\begin{center}
       { {\large\bf
        J. M. Aroca }
	\vskip 0.5cm
        Departament de Matem\`atiques, \\
        Universitat Polit\`ecnica de Catalunya, \\
        Escola T\`ecnica Superior d'Enginyers de Telecomunicaci\'o, 08034 \\
        Barcelona, Spain
        \ \\
\vspace{ 7 mm}
        and
\vspace{ 7 mm}
        \ \\
        {\large\bf  M. Baig and H. Fort}
	\vskip 0.5cm
        Grup de F\'\i sica Te\`orica\\
        and\\
        Institut de F\'\i sica d'Altes Energies\\
        Universitat Aut\`onoma de Barcelona\\
        08193 Bellaterra (Barcelona) Spain}
\end{center}

\vspace{\fill}

\nopagebreak

\begin{abstract}
It is showed how the Hamiltonian lattice
$loop$ $representation$ can be cast
straightforwardly in the Lagrangian formalism. The procedure is
general and here we present the simplest case:
pure compact QED. This connection has been shaded
by the non canonical character of the algebra of the fundamental
loop operators. The loops represent tubes of electric flux
and can be considered the dual objects to the Nielsen-Olesen
strings supported by the Higgs broken phase.
The lattice loop classical action corresponding to the Villain form
is proportional to the
quadratic area of the loop world sheets and thus it is similar to
the Nambu string action.
This loop action is used in a Monte Carlo simulation and its appealing
features are discussed.
\end{abstract}

\end{titlepage}

\section{Introduction}

A unified quantum theory which describes the gauge fields and the
gravitation is one of the main goals pursued by the physicists for long time.
A good candidate for accomplishing this
comprehensive framework is the {\em loop representation}.
This loop approach was introduced
in the early eighties by Gambini and Trias \cite{gt},\cite{gt1} as a
Hamiltonian representation of gauge theories in terms of their natural
physical excitations: the loops.
The original aim of this general analytical Hamiltonian approach
for gauge theories was to avoid the redundancy introduced
by the gauge symmetry working directly in the space of
physical states. However, soon it was realized that the loop
formalism goes far beyond of a simple gauge invariant description.
The introduction by Ashtekar \cite{a} of a new set of variables
that cast general relativity in the same language as
gauge theories allowed to apply loop techniques as a
natural non-perturbative description of Einstein's theory.
Furthermore, the loop representation appeared as the most
appealing application of the loop techniques to this problem
\cite{rs},\cite{gam}.

\vspace{3mm}

The Hamiltonian techniques for gauge
theories have been
developed during the last decade and they provide interesting results
for several lattice models \cite{gt2}-\cite{af}. On the other hand
a Lagrangian approach in terms of loops has been elusive, due mainly
to the non-canonical character of the loop algebra. This feature forbids
the possibility of performing a Legendre transformation as
a straightforward way to obtain the Lagrangian from the Hamiltonian.

In the case of non-Abelian gauge theory a major problem has been
whether we can write a reasonably simple Lagrangian in terms of
"electric vector potentials" \cite{Ma}.
A Lagrangian loop formulation will give rise to new computation
techniques providing a
a useful complement to the Hamiltonian loop studies.

Recently, it was proposed a tentative classical
action in terms of loop variables for the U(1) gauge
theory \cite{aggs}.
Shortly afterwards, we proved that the lattice version of this action
is equivalent to Villain form for D=2+1 dimensions but is slightly
different for D=3+1 dimensions \cite{af1}. In fact, this action
written in terms of variables directly attached to spatial loops
seems to fail in describing all the dynamical degrees of freedom
for D=4.

\vspace{3mm}

This paper is organized as follows. In section 2 we show how the
loops, originally thought up
within the Hamiltonian formalism, can be introduced in a natural
way in the lattice Lagrangian theory. We follow a
different approach to that of reference \cite{af1}: we show
how the electric loops can be traced in the statistical lattice
formulation of $4d$ $U(1)$ theory giving rise to
an expression of the partition function
as a sum of integer closed surfaces.
We interpret these surfaces as the world sheets of the
electric loops.
We discuss the connection of this classical loop action with
the Nambu string action. A clear analogy is patent between the
gauge theory in the loop representation and
the bosonic and fermionic strings, as it was previously suspected
\cite{P}, \cite{gn} but never (as far as we know) demonstrated
explicitly. The parallelism of the
loop representation with the topological representation
of the broken Higgs phase in terms of Nielsen-Olesen strings
\cite{no} is also pointed out.
In section 3 we use the loop action equivalent to the Villain form
for performing a Monte Carlo simulation. It turns out
that this action is the same as the
$\gamma \rightarrow \infty$ limit for the non-compact Abelian-Higgs
theory \cite{bd} ($\gamma$ : Higgs coupling constant).

\vspace{5 mm}

\section{The Lagrangian loop Representation}

The loop based approach of ref.\cite{gt} describes the quantum
electrodynamics in terms of the gauge invariant holonomy
(Wilson loop)
\begin{equation}
\hat{W} (\gamma ) = \exp [i e \oint_{\gamma} A_a (y) dy^a] ,
\label{eq:Wloop}
\end{equation}
and  the  conjugate  electric  field $\hat{E}^a (x)$ .   They  obey   the
commutation relations
\begin{equation}
[\hat{E}^a(x), \hat{W} (\gamma)] = e \int_\gamma \delta(x - y) dy^a \hat{W}
(\gamma) .
\label{eq:alg}
\end{equation}

     These  operators  act  on  a  state  space  of  abelian  loops
$\psi(\gamma)$ that may be expressed in terms of the transform

\begin{equation}
\psi (\gamma) = \int d_\mu [A] <\gamma \mid A> <A \mid \psi>
= \int d_\mu [A] \psi [A] \exp [- i e \oint_{\gamma} A_a dy^a] .
\label{eq:trans}
\end{equation}

This loop representation has many appealing features: In first place,
it allows to do away with the first class constraints of gauge theories.
That is, the Gauss law is automatically satisfied. In second place, the
formalism only involves gauge invariant objects. Finally, all the
gauge invariant operators have a transparent geometrical meaning
when they are realized in the loop space.

When this loop representation is implemented in the lattice it
offers a gauge invariant description
of physical states in terms of kets
$\mid C >=\hat{W}(C) \mid 0>$,
where $C$ labels a closed path in the {\em spatial} lattice.
Eq.(\ref{eq:alg}) becomes
\begin{equation}
[\hat{E}_l,\hat{W}(C)] =  N_l(C)\hat{W}(C),
\label{eq:alg1}
\end{equation}
where $l$ denotes the links of the lattice, $\hat{E}(l)$
the lattice electric field operator,
$\hat{W}(C)=\prod_{l\in C}\hat{U}(l)$ and
$N_l(C)$ is the number of times that the link $l$
appears in the closed path $C$.

In this loop representation, the Wilson loop
acts as the loop creation operator:

\begin{equation}
\hat{W}(C')\mid C> = \mid C'\cdot C>.
\label{eq:Wloop1}
\end{equation}

     The physical meaning of an abelian loop may be deduced  from
(\ref{eq:alg1}) and (\ref{eq:Wloop1}), in fact
\begin{equation}
\hat{E}_l \mid C> = N_l(C) \mid C>,
\label{eq:E}
\end{equation}
which implies  that  $\mid C>$  is  an  eigenstate  of  the
electric field. The corresponding eigenvalue is different from
zero if the link $l$ belongs to
$C$. Thus $C$ represents a confined  line  of  electric flux.

\vspace{3mm}

In order to cast the loop representation in Lagrangian form
it is convenient to use the language of differential
forms on the lattice of ref.\cite{g}.
Besides the great simplifications
to which this formalism lead its advantages
consists in the general character of the expressions obtained.
That is, most of the transformations are independent on the space-time
dimension or on the rank of the fields. So let us sumarize the main
concepts and some useful results of the formalism of differential forms
on the lattice.

	A k-form is a function defined on the k-cells of the lattice
(k=0 sites, k=1 links, k=2 plaquettes, etc.) over an abelian group
which shall be {\bf R}, {\bf Z}, or U(1)={reals module 2$\pi$}.

    Integer forms can be considered geometrical objects on the lattice.
For instance, a 1-form represents a path and the integer value on a link
is the number of times that the path traverses this link.

     Let us introduce
$\nabla$ is the co-border operator which
maps k-forms onto (k+1)-forms. It is the gradient operator when acting
on scalar functions (0-forms) and it is the rotational on vector functions
(1-forms). We shall consider the scalar product of p-forms defined
$<\alpha \mid \beta> = \sum_{c_k}\alpha (c)\beta (c)$ where the sum runs
over the k-cells of the lattice. Under this product the $\nabla$
operator is adjoint to the border operator $\partial$ which maps
k-forms onto (k-1)-forms and which corresponds to minus times the usual
divergence operator. That is,

\begin{eqnarray}
<\alpha \mid  \nabla \beta> = <\partial \alpha \mid \beta>,\\
<\nabla \alpha \mid \beta> = <\alpha \mid \partial \beta>.
\label{eq:inter}
\end{eqnarray}

The co-border $\nabla$ and border $\partial$ operators verify

\begin{equation}
{\nabla}^2 = 0,\;  \;  {\partial}^2 = 0.
\label{eq:0sq}
\end{equation}

The Laplace-Beltrami operator operator is defined by

\begin{equation}
\Box =\nabla \partial +\partial \nabla .
\label{eq:Box}
\end{equation}

It is a symmetric linear operator which commutes with $\nabla$ and $\partial$,
and differs only by a minus sign of the current
Laplacian $ \Delta_\mu \Delta_\mu$.

\vspace{3mm}

{}From Eq.(\ref{eq:Box}) is easy to show the Hodge-identity:

\begin{equation}
1=\partial {\Box}^{-1}\nabla  + \nabla {\Box}^{-1}\partial .
\label{eq:Hodge}
\end{equation}

A useful tool to consider is the {\em duality transformation} which maps
biyectively k-forms
over (D-k)-forms. We denote by $*p_{c_{D-k}}$ the dual form of the
$p_{c_k}$ form. For example, for $D=2$, to plaquettes there correspond
sites of the dual lattice, i.e. those vertices obtained from the original
ones by a translation of vector (a/2, a/2).

Under duality the border and co-border operators
interchange:

\begin{equation}
\partial = *\nabla * .
\label{eq:dualrel}
\end{equation}

\vspace{3mm}

After this digresion about differential forms on the lattice
let us consider the generating functional for the Wilson
U(1) lattice action:
\begin{equation}
Z_W = \int_{-\pi}^{\pi} (d\theta_l )\exp(-\frac{\beta}{2}\sum_p \cos
{\theta}_p),
\end{equation}
where the subscripts $l$ and $p$ stand for the lattice links
and plaquettes respectively.

     Fourier expanding the $\exp [ \cos \theta ]$ we get
\begin{equation}
Z_W = \int_{-\pi}^{\pi} (d\theta_l )  \prod_p \sum_{n_p} I_{n_p}(\beta )
e^{in_p {\theta}_p},
\end{equation}
which can be written, using the language of differential forms as
\begin{equation}
Z_W = \sum_{ \left\{ n_p \right\} }
\int_{-\pi}^{\pi} (d\theta_l ) \exp (\sum_p \ln I_{n_p}(\beta ) )
e^{i<n , \nabla \theta_l >}.
\end{equation}

In the above expression, $\theta_l$ is a real periodic 1-form, that is,
a real number $\theta \in [-\pi, \pi ]$
defined on each link of the lattice; $\nabla$ is the co-border operator;
$n_p$ are integer 2-forms, defined
at the lattice plaquettes.

By eq. (\ref{eq:inter}) and integrating over $\theta_l$
we obtain a $\delta (\partial n_p)$.
Then,
\begin{equation}
Z_W \propto \sum_{ \left\{ n_p; \partial n_p = 0 \right\} }
\exp (\sum_p \ln I_{n_p}(\beta ) ),
\label{eq:Wil}
\end{equation}
the constraint  $\partial n_p = 0$ means that the sum is restricted to
$closed$ 2-forms. Thus, the sum runs over collections of plaquettes
constituting closed surfaces. This expression was obtained by Savit
\cite{s} as an intermediate step towards the $dual$ representation.

An alternative and more easy to handle lattice action than the Wilson
form is the Villain form. The partition function of that form
is given by
\begin{equation}
Z_V = \int (d\theta ) \sum_{ s }\exp
(-\frac{{\beta}_V}{2}\mid\mid \nabla \theta -2\pi s\mid\mid^2),
\label{eq:Villain}
\end{equation}
where $\mid \mid \ldots \mid \mid^2 = <\ldots , \ldots>$.
If we use the Poisson summation formula
$$\sum_s f(s) = \sum_n \int_{-\infty}^{\infty} d\phi
f(\phi) e^{2\pi i\phi n}$$
and we integrate the continuum $\phi$ variables we get
\begin{equation}
Z_V = (2\pi {\beta}_V)^{-N_p/2} \int (d\theta ) \sum_{n }\exp
(-\frac{1}{2{\beta}_V}<n,n>+i<n,\nabla \theta >),
\label{}
\end{equation}
where $N_p$ in the number of plaquettes of the lattice. Again, we can use
the equality:
$<n,\nabla \theta>=<\partial n,\theta >$ and integrating over $\theta$
we obtain a $\delta (\partial n)$. Then,
\begin{equation}
Z_V = (2\pi {\beta}_V)^{-N_p/2}  \sum_{\left\{ n; \partial n = 0 \right\}}
\exp(-\frac{1}{2{\beta}_V}<n,n>),
\label{eq:Vill}
\end{equation}
where $n$ are integer 2-forms. Eq. (\ref{eq:Vill}) is
obtained from Eq.(\ref{eq:Wil}) in the $\beta \rightarrow \infty$
limit.

If we consider the intersection of one of such surfaces
with a $t=constant$ plane we get a loop $C_t$. It is easy
to show that the creation operator of this loop is just
the creation operator of the loop representation, namely
the Wilson loop operator. Repeating the steps from
Eq.(\ref{eq:Villain}) to Eq.(\ref{eq:Vill}) we get for
$<\hat{W}(C_t)>$

\begin{equation}
<W(C_t)> = \frac{1}{Z}(2\pi {\beta}_V)^{-N_p/2}
\sum_{
                \begin{array} {c} n\\
                              \left(  \partial n = C_t \right)
                \end{array}
                 }\exp
(-\frac{1}{2{\beta}_V}<n,n>).
\label{eq:W}
\end{equation}

This is a sum over all closed world sheets and over all world sheets
spanned on the loop $C_t$.
In other
words, we have arrived to an expression of the partition
function of compact electrodynamics in terms of the world
sheets of loops: the $loop$ (Lagrangian) representation.

There are other equivalent representations
which can be obtained from
the Villain form. First, we have the the $dual$
representation \cite{s} obtained
essentially by using the Poisson identity and then performing a
duality transformation. Actually, the loop representation
for the compact U(1) gauge model is reached following this procedure
but stoping before the duality transformation.
Second, for any lattice theory with Abelian
compact variables,
the `$topological$' or $BKT$ (for Berezinskii-Kosterlitz-Thouless)
representation \cite{bwpp} via the `$Banks-Kogut-Myerson$'
transformation \cite{bkm}. The $BKT$ expression for the partition
function of a lattice theory with Abelian compact variables is given by
\begin{equation}
Z_V \propto \sum_{
                \begin{array} {c} *t\\
                              \left(  \partial *t = 0 \right)
                \end{array}
                 }\exp
(-\frac{2\pi^2}{g^2} <*t,\hat{\Delta}*t>),
\end{equation}
i.e. a sum over closed $(D-k-2)$ topological forms $*t$ attached to
the cells $c_{(D-k-2)}$ of the dual lattice and where $\hat{\Delta}$
represents the propagator operator. In the case of compact
electrodynamics, $*t\equiv *m$
i.e. the topological objects are monopoles (particles for D=2+1 and
loops for D=3+1) and $\hat{\Delta}\equiv \frac{1}{\Box}$.

\vspace{3 mm}

	Returning to the loop representation of the partition
function Eq.(\ref{eq:Vill}) we can observe that the loop action
is proportional to the $quadratic$ $area$ $A_2$:
\begin{equation}
S_V = -\frac{1}{{\beta}_V} A_2 =
-\frac{1}{{\beta}_V}\sum_{p \in {\cal S}} n_p^2 = -\frac{1}{{\beta}_V}<n,n>,
\label{eq:A2}
\end{equation}
i.e. the sum of the squares
of the mul\-ti\-pli\-ci\-ties $s_p$ of pla\-que\-ttes which
constitute the loop's
world sheet { \cal S}. It is interesting to note the similarity
of this action with the continuous Nambu action
or its lattice version, the Weingarten action \cite{we}
which are proportional to the area swept out by the bosonic string
\footnote{The relation between the surfaces of the Wilson action and
those of Weingarten action has been analyzed by Kazakov et al
in ref.\cite{kkm}.}.

On the other hand,
we know that in the continuum the classical action of
topological string-like solitons, namely
Nielsen-Olesen vortices, reduces to the Nambu action in the
strong coupling limit \cite{no}.
The Nielsen-Olesen strings are static solutions of the Higgs
Abelian model. A patent analogy is observed when we compare the
$loop$-representation of lattice compact pure $QED$ and the
$BKT$-representation of lattice $Higgs$ non-compact $QED$.
(We compare with the non-compact instead of the compact version
because this last, in addition to Nielsen-Olesen strings, also has
Dirac monopoles as topological solutions and then we have to
consider open as much as closed world sheets. In ref.
\cite{af2} we considered the {\em Higgs compact QED} which exhibits
duality between both representations).

In the case of $Higgs$ non-compact $QED$ model
( a non compact gauge field $A_\mu$
coupled to a scalar field $\Phi=|\Phi|e^{i\phi}$) we have
$t*\equiv \sigma*$, where $\sigma*$ represents a 2-form
which corresponds to the world sheet of the topological objects
namely the Nielsen-Olesen strings
\cite{pwz} and $\hat{\Delta}\equiv \frac{1}{\Box+M^{2}}$ (M is the
the mass acquired by the gauge field due to the Higgs mechanism).
Thus, both models consist in a sum over closed surfaces
which are the world sheets of closed electric
strings (loops) and closed
magnetic strings (closed Nielsen-Olesen vortices) respectively.
The corresponding lattice actions are essentially the quadratic area
of the world sheets in both cases.
Moreover, the creation operator of both loops and N.O. strings
is essentially the same: the Wilson loop operator \cite{gt} \cite{pwz}.

It is also possible to regard the connection between this two
models from a different point of view : the Villain form of
$U(1)$ is the Higgs coupling
$\rightarrow \infty$ limit of {\em Higgs non-compact QED}.
The standard action of {\em Higgs non-compact QED} is given by
\begin{equation}
S = -\frac{\beta}{2} \sum_p {\theta}_p^2 + \gamma \sum_l
\bar{\phi}_xU(l)\phi_{x+l}.
\label{eq:sncqed}
\end{equation}

It was pointed out in reference \cite{bd} that in the limit
$\gamma \rightarrow \infty$
$$U(l) = e^{i{\theta}_l} \rightarrow 1,$$
which implies for the angular variables
$${\theta}_l=2\pi n_l,$$
and so, the action (\ref{eq:sncqed}) becomes
\begin{equation}
S_\infty = -\frac{\beta}{2}\sum_{p \in {\cal S}} (2\pi n_p)^2 =
-2{\pi}^2\beta<n,n>,
\end{equation}
where $n_p=\nabla n_l$. We note that this is just the Villain action
of gauge $U(1)$ model, Eq.(\ref{eq:A2}) but with $2{\pi}^2\beta$
instead of $\frac{1}{{\beta}_V}$.

\vspace{1cm}

\section{Numerical Computations}
\bigskip

	Here we shall present the results of numerical simulations
carried out for the loop action (\ref{eq:A2}) corresponding
to Villain form of $U(1)$ model.
In fact, we have simulated the $dual$ action in terms of the
dual integer variables $*n_p$
\begin{equation}
S_d = -\frac{1}{{\beta}_V}\sum_{p} *n_p^2 = -\frac{1}{{\beta}_V}
\sum_{p} (\nabla *n_l)^2
= -\frac{1}{{\beta}_V} \sum_{p} (\sum_{l \in p} *n_l)^2  ,
\label{eq:dual}
\end{equation}
where $*n_l$ are integer 1-forms attached to links of the dual lattice
and the integer 2-forms $*n_p$ correspond to their lattice curl, i.e.
$\partial n_p = 0$ means that $n_p=\partial n_c$ where $n_c$ are integer
3-forms attached to the elementary cubes and according to (\ref{eq:dualrel})
$*n_p=\nabla *(n_c) = \nabla *n_l$.
Note that this action implies the assignment of unbounded integer
variables to the links of the lattice and the action is defined through
the square of the ordered sum of the integers of an elementary square of
the lattice. We have implemented a  Metropolis algorithm fixing
the acceptance ratio, as it is usual in random surfaces analysis.

We have studied this model for different
lattice sizes, imposing periodic boundary conditions. Simple thermal
cycles showed the presence of a phase transition in the
neighborhood of $\beta_V=0.639$. To
study the order of this transitions we have analyzed the energy
histogram and we have checked for the presence of tunneling between the
phases. We have not applied any reweighting extrapolation technique,
only a direct observation of the histograms.

In Fig. 1a we present the histogram of the plaquette energy density
including 80.000 measures, after discarding 40.000 thermalizing
iterations,
for a $12^4$ lattice just at the Villain
transition point ${\beta}_V = 0.639$.
One can observe clearly a two-peak structure.

In Fig. 1b we present the
time evolution of the total internal energy during the simulation. Each
point is the average over 100 consecutive iterations.
This analysis shows clearly the
presence of "tunneling" between the phases.

Our numerical results, using this loop action equivalent to Villain form
and imposing periodic boundary conditions ($PBC$), exhibit a
first order phase transition signal.
This is
in agreement with the standard lattice numerical simulations of
QED using the Wilson action
(again using the standard $PBC$) \cite{jnz},\cite{acg}.
Nevertheless, some differences between the simulations using
 Wilson or  Villain actions
arise. In particular, we have not observed here the strange
persistence of the phases -and the absence of tunneling- seen in the
simulations performed with the Wilson action. Remember that
according \cite{lne}-\cite{bf} the first order
nature of this phase transition seems to be  a spurious effect
produced by the non-trivial topology that periodic boundary
conditions bear with.

Typical loop configurations, obtained by intersecting the
lattice with $x_4$ = constant planes, are showed in Fig.2 for
two values of the coupling constant, one to the left (strong coupling)
and one to the right (weak coupling) of the transition point.
The difference between the two typical configurations is patent.
These
configurations are obtained by taking the last measure after thermalization
with 20.000 iterations and they
represent all the plaquettes with
integer value $n_p \ne 0$.
In order to obtain this plaquettes we proceed as follows:
first, we stored the $*n_p \neq 0$ of the
dual plaquettes and then we performed a duality transformation to
get the corresponding non zero $n_p$.

In TABLE 1 we present the spectra of plaquettes $\left\{ n_p \right\}$
(in all the lattice, not only in
a particular cut $x_4 = t$ ) for different
values of the ${\beta}_V$ coupling.
Comparing the $\left\{ n_p \right\}$ at ${\beta}_V$=0.637 and
${\beta}_V$=0.641 i.e. just before and after the
critical Villain coupling $\beta_c$=0.639 one can observe
an abrupt increment of non zero plaquettes.

\begin{center}
TABLE 1.
\end{center}
\vspace{0.5 cm}

\nopagebreak

\begin{tabular}{||c|l|l|l|l|l|l|l|l|l||}
\multicolumn{10}{c}{Plaquette Configurations}        \\ \hline
${\beta}_V$ &$n=-4$ &$n=-3$ &$n=-2$ &$n=-1$ & $n=0$  & $n=1$ & $n=2$ &
$n=3$ & $n=4$ \\ \hline
   0.200    &  0    &   0   &    0  &    15 & 124386 &    15 &    0  &
0  &   0   \\ \hline
   0.500    &  0    &   0   &    2  &  1657 & 121099 &  1659 &    0  &
0   &  0   \\ \hline
   0.555    &  0    &   0   &    2  &  3661 & 117093 &  3655 &    5  &
0   &  0   \\ \hline
   0.637    &  0    &   0   &   70  & 12208 &  99850 & 12228 &   60  &
0   &  0   \\ \hline
   0.641    &  0    &   0   &  115  & 15104 &  93972 & 15116 &  109  &
0   &  0   \\ \hline
   0.714    &  0     &   0  & 299   & 19573 &  84639 & 19639 & 266   &
0   &  0   \\ \hline
   1.000    &  1     &  12  & 1292  & 25787 &  70239 & 25776 & 1300  &
9   &  0   \\ \hline
   2.000    & 12     & 569  & 6695  & 30141 &  49474 & 30337 & 6630  &
543 & 15   \\ \hline
\hline
\end{tabular}

\vspace{0.5 cm}

Incidentally, we want to remark that
the apparent first order behaviour of the
{\em Higgs non-compact QED}
in the $\gamma \rightarrow \infty$ limit
found in ref. \cite{bd}
can be clarified under this new perspective.
In that limit it turns out a sort of "compactification"
which transforms the original model in the pure {\em compact QED}
(Villain form) we have studied here.

\section{Conclusions}

As it was mentioned, the loop space provides a common scenario for a
non-local description of gauge theories and quantum gravity.
Up to now, the loop approach was exclusively a Hamiltonian
formalism and no lagrangian counterpart was available. A classical
action for the Yang-Mills theory in terms of loop variables would be very
valuable in its own right because they are the natural
candidates to describe the theory in a confining phase.
In addition, it may be useful to obtain semiclassical approximations
to gauge theories or to general relativity in terms of
Ashtekar's variables. Here we present some small steps
in this direction which continue those ones of ref. \cite{aggs}
and \cite{af1}.

\vspace{3mm}

In relation with the considered case of lattice Abelian gauge theory,
we observed the known analogy between the confining
phase of lattice electrodynamics described in terms of electric loops
and the Higgs phase described in terms of Nielsen-Olesen magnetic
vortices.

We can also
ask whether the loops are no more than a useful
representation or if, perhaps, they have a deeper physical meaning.
Lattice QED exhibits a confining-deconfining transition, although
in principle, ordinary continuum QED has only one non-confining phase.
However, there are studies which indicate that also there is
a phase transition for QED in the continuum \cite{mn}. In addition to
the usual weak coupling phase, a strong coupling confining phase
exists above a critical coupling $\alpha_c$. This new phase
could explain a mysterious collection of data from heavy ion
collisions \cite{sch}-\cite{cg}.
The unexpected feature is the observation of positron-electron
resonances with narrow peak energy in the range of 1.4-1.8 MeV.
This suggest the existence of 'electro-mesons' in a strongly coupled
phase of QED. Moreover, a new two-phases model of continuum QED
and a mass formula for taking account of the
positronium spectrum in the strong coupling phase has been regarded
recently \cite{az}.
Thus, in principle, we can speculate about the existence in nature of
abelian electric tubes providing a real support for abelian loops and
the possibility of
being on equal footing with the observed magnetic vortices.

\vspace{3mm}
{\large \bf Acknowledgements}
\vspace{3mm}

This work has been supported in part by the CICYT
project AEN93-0474, H.F. has
been supported by the COMMISSION OF THE EUROPEAN COMMUNITIES.

\vspace{2mm}

We wish to thank Rodolfo Gambini for
useful discussions and comments.

\newpage

{\large \bf Figure Captions.}

\vspace{0.5cm}

\begin{description}

\item[1a] Histogram of the plaquette energy density
corresponding to 80.000 measures on a $12^4$ lattice just at the Villain
transition point ${\beta}_V = 0.639$.

\item[1b] Time evolution of the total internal energy during the simulation.
Each point is the average over 100 consecutive iterations.

\item[2a] Typical loop configurations, obtained by intersecting the
lattice with $x_4=6$, at ${\beta}_V =0.5$ (strong coupling phase).

\item[2b] Typical loop configurations, obtained by intersecting the
lattice with $x_4=6$, at ${\beta}_V =1$ (weak coupling phase).

\end{description}


\newpage



\begin{thebibliography}{99}

\bibitem{gt} R.Gambini and A.Trias, Phys.Rev.D {\bf 22} (1980) 1380.

\bibitem{gt1} R.Gambini and A.Trias, Phys.Rev.D {\bf 23} (1981) 553.

\bibitem{a}  A. Ashtekar { Phys. Rev. Lett.} {\bf 57}, (1986) 2244;
 A. Ashtekar, { Phys. Rev.} {\bf D36}, (1987) 1587.


\bibitem{rs}  C. Rovelli and L. Smolin, { Phys. Rev. Lett.} {\bf 61}
(1988) 1155; { Nuc. Phys.} {\bf B331} (1990) 80.

\bibitem{gam} R. Gambini { Phys. Lett.} {\bf B235} (1991) 180;

\bibitem{gt2} R.Gambini and A.Trias, Nuc. Phys. {\bf B278} (1986) 436.

\bibitem{glt} R.Gambini, L.Leal and A.Trias, Phys.Rev.{\bf D39} (1989)
              3127.
\bibitem{dgt} C. Di Bartolo, R.Gambini and A.Trias, Phys.Rev.{\bf D39}
              (1989) 1756.

\bibitem{af} J.M. Aroca and H. Fort, Phys.Lett.{\bf B299} (1993) 305;
             J.M. Aroca and H. Fort, Phys.Lett.{\bf B317} (1993) 604.



\bibitem{Ma} S. Mandelstam, Phys. Rev. {\bf D19} (1979) 2391.

\bibitem{aggs} D.Armand Ugon, R.Gambini, J.Griego and L.Setaro,
                                Preprint IF/FCN-93 July, 1993.

\bibitem{af1} J.M.Aroca and H.Fort, Phys.Lett.{\bf B325} (1994) 166.

\bibitem{P} A. M. Polyakov, { Nuc. Phys.} {\bf B164}, (1979) 171.

\bibitem{gn} J.L. Gervais and A. Neveu, { Phys. Lett.}{\bf 80B} (1979) 255.

\bibitem{no}   H.B.Nielsen and P.Olesen, Nucl. Phys.
                      {\bf B61} (1973) 45.

\bibitem{bd}   M. Baig and E. Dagotto, Nucl. Phys. {\bf B17} (Proc. Suppl.)
               (1990) 671; M. Baig, E. Dagotto and E. Moreo Phys. Lett.
               {\bf 242} (1990) 444.

\bibitem{g} A. H. Guth, Phys.Rev.{\bf D21} (1980) 2291.

\bibitem{s}    R.Savit, Rev.Mod.Phys.{\bf 52} (1980) 453.

\bibitem{bwpp} A.K. Bukenov, U.J. Wiese, M.I.Polikarpov and A.V. Pochinskii
              , Phys.At.Nucl.{\bf 56} (1993) 122.

\bibitem{bkm}  T. Banks, R. Myerson and J.B. Kogut, Nucl. Phys.
                      {\bf B129} (1977) 493.
\bibitem{we} D. Weingarten, Phys.Lett. {\bf B90} (1980) 280.

\bibitem{kkm} V.A. Kazakov, T.A. Kozhamkulov and A.A. Migdal,
              Sov.J.Nucl.Phys. {\bf 43} (1986) 301.

\bibitem{af2} J.M.Aroca and H. Fort Preprint UAB-FT-329/94.

\bibitem{pwz} M.I.Polikarpov U.J.Wiese and M.A.Zubkov, Phys.Lett.
               , {\bf B309} (1993) 133.


\bibitem{jnz}  J. Jerzak, T. neuhaus and P.M. Zerwas, Phys.Lett. {\bf
               B133} (1983) 103.

\bibitem{acg} V. Azcoiti, G. Di Carlo and A. F. Grillo, Phys. Lett.
		{\bf B267} (1991) 101.

\bibitem{lne} C.B. Lang and T. Neuhaus, Nuclear Physics {\bf B} (Proc.
              Suppl.) {\bf 34} (1994) 543.

\bibitem{rebbi} W. Kerler, C. Rebbi and A. Weber, Preprint BU-HEP 94-7.

\bibitem{bf}    M. Baig and H. Fort, Preprint UAB-FT 338/94 (Phys.
Lett. B, in press)

\bibitem{mn}  T. Maskawa and H. Nakajima, Progr.Theor.Phys. {\bf 52}
         (1974) 1326; T. Maskawa and H. Nakajima, Progr.Theor.Phys.{\bf 54}
              (1975) 860;
R.Fukuda and T.Kugo, Nucl.Phys.{\bf B117} (1976) 250;
V.A.Miransky, Nuovo Cimento {\bf 90A} (1985) 149;
V.A.Miransky and P.I.Fomin, Sov.J.Part.Nucl.{\bf 16} (1985) 203;

\bibitem{sch} J. Schweppe et al, Phys.Rev.Lett.{\bf 51} (1983) 2261.

\bibitem{c} M.Clemente et al, Phys.Lett.{\bf B137} (1984) 41.

\bibitem{co} T. Cowan et al, Phys.Rev.Lett.{\bf 54} (1985) 1761;
T. Cowan et al, Phys.Rev.Lett.{\bf 56} (1986) 444.

\bibitem{t} H. Tsertos et al, Phys.Lett.{\bf B162} (1985) 273;
 Z.Phys.{\bf A326} (1987) 235.

\bibitem{cg} T. Cowan and J.Greenberg,in {\it Physics of Strong Fields}
edited by W.Greiner (Plenum, New York, 1987).

\bibitem{az}  Awada and Zoller,  Nucl. Phys. {\bf B365} 699 (1993).


\end{thebibliography}
\end{document}